# High-speed KATAN Ciphers on-a-Chip

Fatma M. Qatan and Issam W. Damaj

Computer Engineering Department
American university of Kuwait, Kuwait
{S00013219, idamaj}@auk.edu.kw

*Abstract*—**Security in embedded systems has become a main requirement in modern electronic devices. The demand for low-cost and highly secure cryptographic algorithms is increasingly growing in fields such as mobile telecommunications, handheld devices, etc. In this paper, we analyze and evaluate the development of cheap and relatively fast hardware implementations of the KATAN family of block ciphers. KATAN is a family of six hardware oriented block ciphers. All KATAN ciphers share an 80-bit key and have 32, 48, or 64-bit blocks. We use VHDL under Altera Quartus in conjunction with ModelSim to implement and analyze our hardware designs. The developed designs are mapped onto high-performance Field Programmable Gate Arrays. We compare our findings with similar hardware implementations and C software versions of the algorithms. The performance analysis of the C implementations is done using Intel Vtune Amplifier running on Dell precision T7500 with its dual quad-core Xeon processor and 24 GB of RAM. The obtained results show better performance when compared with existing hardware and software implementations.**

I. INTRODUCTION

Nowadays, securing information is critically important, especially when it comes to portable devices such as tablets, mobile phones, etc. In addition to the required high-levels of security and performance, developing low-power and small circuits is equally desirable. However, a great deal of assistance in creating low-power and high-speed cores comes from the inherent simplicity of the selected algorithm for embedding as a hardware component.

Cryptographic algorithms are widely used to ensure confidentiality and integrity of information in various applications. Tiny or lightweight block ciphers are employed for security in environments where resources are limited. Many block ciphers are proposed in the literature and they largely differ in terms of size and performance. Skipjack [1], HIGHT [2], XTEA [3], KATAN [4], PRESENT [5], MCrypton [6], SEA [7], and CGEN [8] are of the many existing lightweight cryptographic algorithms and are efficiently employed for security in environments running on low resources. The structures of tiny ciphers are sufficiently strong, which makes the algorithms safe enough and a good choice for security solutions on most machines.

KATAN is one of the existing lightweight families of block ciphers. The KATAN family splits into two sets. The first set is the KATAN and takes blocks of 32, 48, or 64 bits. The second set is the KTANTAN; it also takes blocks of 32, 48, or 64 bits but differs in the key scheduler. All ciphers in the KATAN family have 80-bit keys.

In addition to being small, less power hungry, and fast, the modifiability, upgradeability, and reusability of security hardware cores are of no less importance. Accordingly, we target field programmable gate arrays (FPGAs) to enable on-the-fly modifications, tuning, and upgrades of the developed designs and implementations.

Nowadays, FPGAs are the cornerstone components of reconfigurable systems. Sizes of FPGAs have increased dramatically in recent years. Companies like Xilinx [9] and Altera [10] produce FPGAs with several millions of gates, such as, Virtex Pro and Stratix FPGAs. Programmable FPGAs, together with modern co-design tools and methodologies, form a powerful paradigm for computing. VHDL and Verilog are two famous hardware description languages (HDLs) that are usually used for implementations using FPGAs.

In this paper, we present the design and implementation of several high-speed and cheap hardware implementations for the KATAN family of block ciphers. The development starts by modeling the designs using a hybrid model that combines flowcharts and concurrent process models. The developed cores are then critically analyzed, evaluated, and benchmarked against similar implementations. The hardware cores are analyzed for their execution time, maximum frequency, propagation delay, throughput and logic area. The targeted hardware system is Altera's Stratix II FPGA. Results are compared to similar implementations in the literature and also to software implementations. The software implementations are analyzed for their execution time and the throughput using the analysis tool Intel Vtune Amplifier. The targeted system is the Dell precision T7500 with its dual quad-core Xeon processor and 24 GB of RAM.

The paper is organized so that Section 2 describes the targeted algorithms. In Section 3, we detail the proposed hardware developments. Section 4 presents the analysis, evaluation, and the comparisons with similar implementations from the literature. Section 5 concludes the paper and plots future directions.

## II. THE KATAN FAMILY OF BLOCK CIPHERS

The KATAN family of block ciphers is notable for its simplicity and conciseness of description. The ciphers were initially presented by De Cannièreet. al in 2009 [4]. The KATAN and KTANTAN sets of ciphers support block sizes of 32, 48, or 64 bits. All ciphers in the KATAN family have 80-bit keys. The main difference between KATAN and KTANTAN is their key schedulers.

KATAN ciphers follow the design of KeeLoq [11]. The plaintext is initially stored in two registers. During each round, several bits are taken from the registers and enter two nonlinear Boolean functions. The output of the Boolean functions is loaded to the least significant bits of the registers after they are shifted. Rounds are executed 254 times to insure sufficient mixing. The structure of a KATAN/KTANTAN round is shown in Fig. 1. The KATAN family is found to be secure against differential and linear attacks.

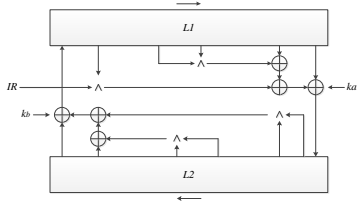

Fig.1: The structure of a KATAN/KTANTAN round showing the registers $L_1$ and $L_2$, in addition to, the subkeys ka and kb and the irregular update IR.

Different hardware implementations for the KATAN family are presented in [4]. The authors presented several results for different design trade-offs. The highest reported speed is around 75 Kbps for both the KATAN and KTANTAN at a frequency of 100 MHz. The results will be discussed in details in Section 4.

## III. THE DEVELOPMENT OF PARALLEL KATAN CIPHERS

The development starts by modelling the system using a hybrid model that combines flowcharts and concurrent process models (CPMs). Flowcharts help in describing the sequential behaviour of the algorithm. The CPM reveals the parallel behaviour of the algorithm. Parallel designs are then captured using VHDL under Quartus. The used development methodology is informal, easy to use, clearly describes the algorithm, and enables smooth capturing of the model under VHDL.

Two different design alternatives are presented in this paper. The first design set relies on the synthesizer to produce parallel implementations starting from behavioural descriptions. The second design set decomposes the parallel implementations into semi-structural pipelines.

The encryption in KATAN ciphers starts by loading the plaintext into the registers L1 and L2. The length of these two registers depends on the size of the plaintext. In the case of KATAN-32, the plaintext consists of 32 bits. L1 and L2 are 13 and 19-bit registers. KATAN-32 uses two nonlinear functions $f_a$ and $f_b$ in each round; the functions are illustrated as follows:

- fa(L1) = L1(x1) ^ L1(x2)^ (L1(x3) · L1(x4)) ^ (L1(x5) · IR) ^ ka
- fb(L2) = L2(y1) ^ L2(y2)^ (L2(y3) · L2(y4)) ^ (L2(y5) · L2(y6)) ^ kb

The encryption processes of the other KATAN ciphers execute similarly but have different block and register sizes.

The structure of KATAN ciphers enables the parallelization of several segments. Few segments can run in a pleasantly parallel fashion. The overall structure can be decomposed into a pipeline (See Figures 2,3,4, and 5).

The encryption method will be decomposed into three main pipelined stages. The first stage consists of three loops that initialize the plaintext and loads the key. The three loops can run concurrently as depicted in Fig. 3.

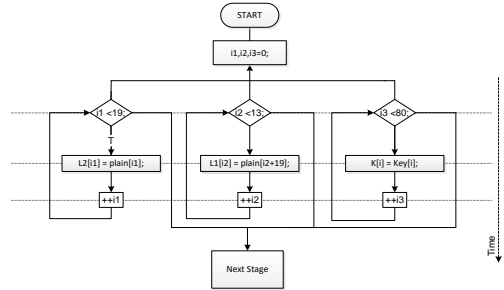

Fig. 3: Plaintext and key initialization; the Initialization stage.

The second stage contains one loop for key scheduling, and one outer loop that has the two nonlinear functions and additional two nested loops. The key scheduler and round stages are depicted in Fig. 4.

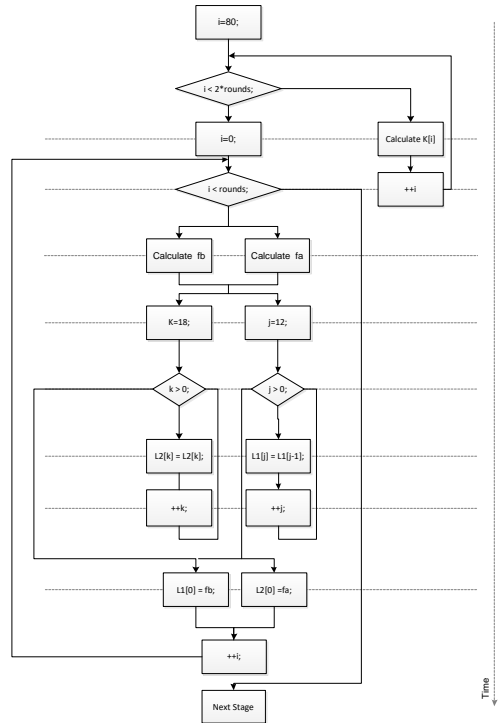

Fig. 4: The Key Scheduler and Round stage.

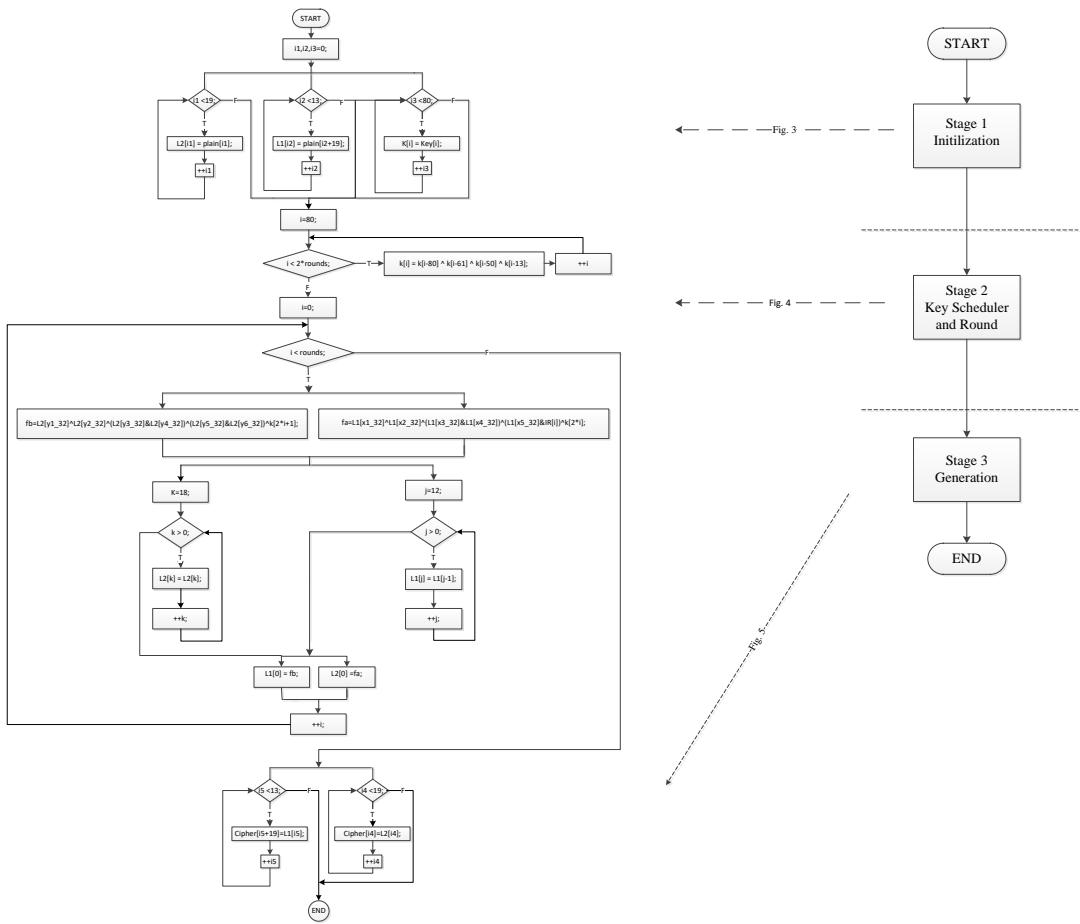

(a)

(b)

U0: KATAN32_1 port map (clk, reset, plain, key1, key2, key3, L11, L22, kk);
U1: reg port map (clk, reset, load, L11, L11Reg);
U2: reg2 port map (clk, reset, load, L22, L22Reg);
U3: reg4 port map (clk, reset, load, kk, kkReg);
U4: KATAN32_2 port map (clk, reset, L11Reg, L22Reg, L111, L222, kkReg);
U5: reg port map (clk, reset, load, L111, L111Reg);
U6: reg2 port map (clk, reset, load, L222, L222Reg);
U7: KATAN32_3 port map (clk, reset, L111Reg, L222Reg, cipher);

(c)

Fig. 2: (a) The hybrid model that combines a flowchart and a concurrent process model for the encryption method of KATAN-32. (b) The decomposed pipeline. (c) Part of the pipelined VHDL implementation.

Table 1: Hardware results of KATAN/KTANTAN block ciphers (Behavioural model), the speedup is for the KTANTAN over the KATAN with the same block size.

| Algorithm Name | Hardware Performance | | | | | | | |
|---|---|---|---|---|---|---|---|---|
| | *Propagation Delay (µs)* | *Number of clock cycles* | *Execution Time (µs)* | *Throughput (Mbits/s)* | *Max Frequency (MHz)* | *ALUTs* | *Logic Elements (LEs)* | *Speedup* |
| KATAN-32 | 0.042 | 35 | 1.470 | 21.769 | 23.970 | 2145 | 3120 | - |
| KATAN-48 | 0.054 | 35 | 1.890 | 25.397 | 18.570 | 3982 | 5775 | - |
| KATAN-64 | 0.068 | 35 | 2.380 | 26.891 | 14.710 | 4315 | 6552 | - |
| KTANTAN-32 | 0.043 | 2 | 0.086 | 372.093 | 21.150 | 1947 | 2808 | 17.09 |
| KTANTAN-48 | 0.056 | 2 | 0.100 | 480 | 16.880 | 3662 | 5460 | 18.89 |
| KTANTAN-64 | 0.073 | 2 | 0.146 | 438.356 | 12.610 | 4075 | 6240 | 16.3 |

Table 2: Hardware results of KATAN/KTANTAN block ciphers (Pipelined model).

| Algorithm Name | Hardware Performance | | | | | | |
|---|---|---|---|---|---|---|---|
| | *Propagation Delay (µs)* | *Number of clock cycles* | *Execution Time (µs)* | *Throughput(Mbits/s)* | *Max Frequency (MHz)* | *ALUTs* | *Logic Elements (LEs)* |
| KATAN-32 | 0.041 | 3 | 0.103 | 312.195 | 24.190 | 2649 | 3900 |
| KATAN-48 | 0.054 | 3 | 0.135 | 355.556 | 18.670 | 4170 | 5928 |
| KATAN-64 | 0.060 | 3 | 0.150 | 426.667 | 16.710 | 3859 | 5772 |

The third stage is composed of two concurrent loops which generate the ciphertext. The third Generation stage is shown in Fig. 5.

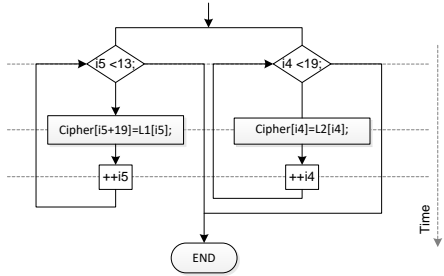

Fig. 5: The ciphertext Generation stage.

IV. RESULTS AND EVALUATION

The performance analysis of the developed designs is done using different tools. The hardware implementations are analyzed using Altera Quartus in conjunction with ModelSim. The obtained results are for the following metrics:

- Propagation delay: the required time for a signal to propagate from an input pin through combinational logic to an output pin.
- The maximum frequency: indicates the clock speed that a certain core is running at.
- Number of clock cycles: the total number of cycles needed to finish execution.
- Execution time: is the overall time that the program takes in order to finish execution.
- Throughput: number of bits encrypted over time; it indicates the speed of the encryption process.
- Chip-area: is the amount of logic occupied by an algorithm mapping onto an FPGA in terms of logic elements (LEs) and adaptive look up tables (ALUTs).

We present three different implementations for the KATAN family. The first implementation is follows the original version presented by the authors in [4]. The second implementation is a behavioural version. The behavioural implementation is decomposed into three different stages to form the third implementation which is the pipelined implementation.

The hardware results for the behavioural designs are shown in Table 1. Among the KATAN implementations, the 32-bit version achieved the smallest chip-area of 2145 ALUTs and 3120 LEs, and the highest operating frequency of around 24 MHz. The fastest KATAN cipher is the 64-bit version with a speed of around 27 Mbits/s. Among the KTANTAN implementations, also the 32-bit version achieved the smallest chip-area of 1947 ALUTs and 2808 LEs, and the highest operating frequency of around 21 MHz. The fastest KTANTAN cipher is the 48-bit version with speed of 480 Mbits/s. Our implementations have achieved speedups up to 1741.5 times over the original implementations reported in [4]. The fastest KTANTAN implementation we have achieved is 16.3 times faster than that the fastest of our KATANs. In Table 1, we draw a comparison between the performance of KATAN and KTANTAN including the speedups.

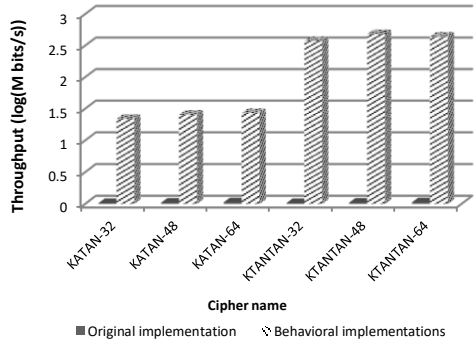

Fig. 6: Throughput of the original implementation vs. our behavioural implementations.

The comparison between our behavioural implementations and the original implementation from [4] is shown in Fig. 6. The performance results show better performance as compared to the original implementations. The KTANTAN-64 cipher achieved the highest throughput of 438.356 Mbits/s while originally had a speed of 25.100 Kbits/s.

The obtained results for the pipelined implementations for the KATAN block ciphers show better performance than the behavioural implementations. The pipelined KATAN-64 cipher achieved the highest throughput of 426.667 Mbits/s while in the behavioural design it achieved a throughput of 26.891 Mbits/s. The comparisons among the behavioural, pipelined, and original implementations are shown in Fig. 7 and Table 3.

As expected, the chip-area of the pipelined 32-bit version is found to be the smallest, among the other pipelined versions, but larger than that of the behavioural version. The pipelined 32-bit version used 2649 ALUTs and occupied 3900 LEs. The structure of the pipelined version is shown in Fig. 8; the entities $U_0$, $U_4$, and $U_7$ are the pipeline stages, while the remaining entities are the buffer registers between the stages.

Table 3: comparison between the original, behavioural and pipelined implementations, $speedup_1$ is for the behavioural over the original while $speedup_2$ is for the pipelined over the original.

| KATAN | Throughput (Mbits/s) | | | $Speedup_1$ | $Speedup_2$ |
|---|---|---|---|---|---|
| | Org. | Beh. | Pip. | | |
| 32-bit | 0.012 | 21.76 | 312.19 | 1741.5 | 24975 |
| 48-bit | 0.018 | 25.39 | 355.55 | 1350.9 | 18912 |
| 64-bit | 0.025 | 26.89 | 426.66 | 1071.3 | 16998 |

We also draw a comparison of the execution time between our hardware results and software versions written in C. The C implementations are compiled and analyzed using Intel Vtune Amplifier, and running on a dual quad-core Xeon processor and 24 GB of RAM. The obtained results are shown in Table 4.

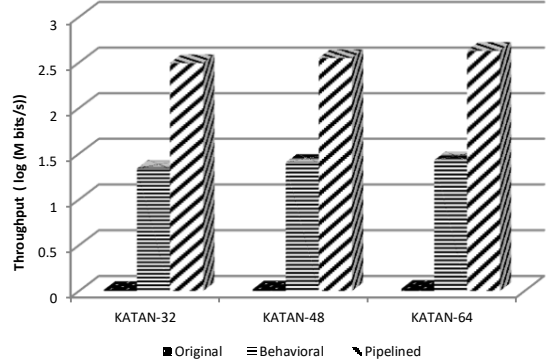

Fig. 7: Throughput of the original, behavioural, and pipelined implementations.

Table 4: Software results of KATAN/KTANTAN block ciphers, $speedup_1$ is for the behavioural over the software implementation, while $speedup_2$ is for the pipelined over software implementation.

| Algorithm Name | Software Performance | | Hardware Performance | |
|---|---|---|---|---|
| | Execution Time (µs) | Throughput (Mbits/s) | Behavioural $Speedup_1$ | Pipelined $Speedup_2$ |
| KATAN-32 | 27.460 | 1.165 | 18.68 | 267.9 |
| KATAN-48 | 40.330 | 1.190 | 21.34 | 298.7 |
| KATAN-64 | 52.830 | 1.211 | 22.21 | 352.3 |
| KTANTAN-32 | 791.080 | 0.040 | 9302.325 | - |
| KTANTAN-48 | 803.320 | 0.060 | 8000 | - |
| KTANTAN-64 | 821.830 | 0.078 | 5619 | |

The improved performance results in the presented implementations are the outcome of several factors, such as, the advances in the synthesizer, the use of a newer technology, and the pipelined structure. The original designs reported in [4] are synthesized with Synopsys Design Vision version Y-2006.06, using UMC 0.13µm Low-Leakage CMOS library. The authors informally reported in [12] new results with higher speed than the results reported in [4] as shown in Table 5. Their new optimized implementations have achieved speedups up to 85712.7 times over the results reported in [4].

Table 5: comparison between the original and new implementations, speedup is for the org.2 over the org.1.

| Algorithm Name | Throughput (Mbits/s) | | Speedup |
|---|---|---|---|
| | Org.1 | Org.2 | |
| KATAN-32 | 0.0125 | 1071.4 | 85712 |
| KATAN-48 | 0.0188 | 1611.4 | 85712.7 |
| KATAN-64 | 0.0251 | 1882.5 | 75000 |
| KTANTAN-32 | 0.0125 | 468.7 | 37496 |
| KTANTAN-48 | 0.0188 | 696.3 | 37037 |
| KTANTAN-64 | 0.0251 | 896.4 | 35713 |

## V. CONCLUSION

The paper presents hardware implementations for the KATAN family of block ciphers. Several behavioural and pipelined designs are developed and mapped onto high-performance FPGAs. The analysis shows an achieved performance higher than the original implementations reported in [4]. The developed hardware cores also outperform software implementations under a powerful high-performance computer. A speedup of around 25k is achieved for the KATAN-32 in the pipelined implementation. The authors informally reported higher speed results than their original implementations; a speedup of around 85.7k is reported. The behavioural KTANTAN-32 achieved the smallest chip-area of 1947 ALUTs and 2808 LEs. Future works include further optimizing the KATAN cores to achieve higher throughputs and/or smaller chip areas.

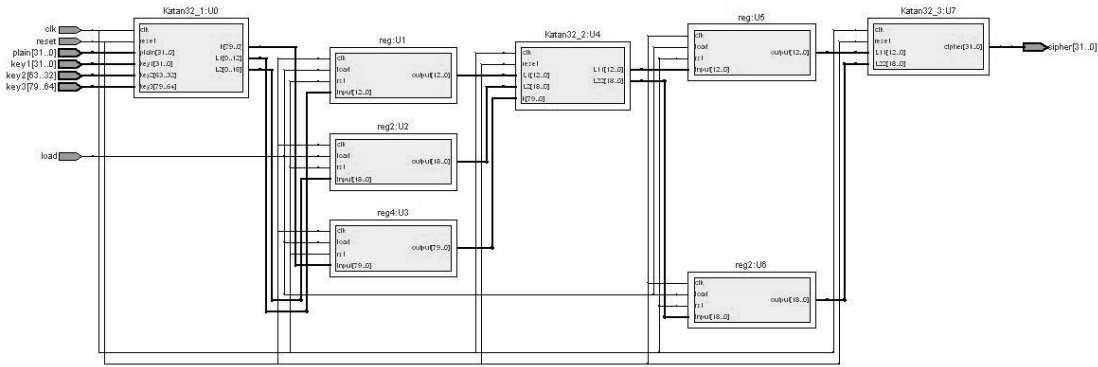

Fig. 8: RTL Viewer of the pipelined KATAN-32.